# Non-linear Cherenkov Muon Spectrometer Using Multi-Layer Pressurized C$_3$F$_8$ Gas Radiators

J. Bae,* S. Chatzidakis*

*School of Nuclear Engineering, Purdue University, West Lafayette, IN 47906, Bae43@purdue.edu*

## INTRODUCTION

In many engineering applications, cosmic ray muons have emerged as a promising radiographic probe for imaging large and dense objects such as spent nuclear fuels[1,2], nuclear reactor cores[3,4], and well-shielded special nuclear materials[5,6] because they are high-energetic, penetrative, and charged leptons. Although cosmic ray muons have obvious advantages over conventional induced radiation probes, i.e., x-rays, their broad application in the nuclear industry is often limited due to their naturally low flux. The cosmic ray muon flux depends on various conditions such as altitude, zenith angle, solar activity, and detector configuration[7–10]. However, it is often simply approximated to $10^4$ m$^{-2}$ min$^{-1}$ at sea level[11]. One of the first successful muon scattering angle trackers was developed by LANL[12]. To reconstruct images of target objects using muon scattering tomography, the scattering angle and momentum of each muon must be measured. In field applications, muon scattering angles are easily reconstructed using the multiple Coulomb scattering approximation, however, it is challenging to measure muon momentum. Therefore, a mean muon momentum, 3 or 4 GeV/c, is often chosen to represent the cosmic ray muon spectrum.

The Cherenkov radiation is often exploited in particle physics for tagging charged particles. For a radiator in a Cherenkov detector, pressurized gas is often preferred because a condition to induce the Cherenkov effect can be manipulated by pressurizing the gas medium. Using the fact that the index of refraction of gas varies with pressure and temperature, a fieldable Cherenkov muon spectrometer based on pressurized CO$_2$ gas was developed in recent studies[13,14]. The CO$_2$ based Cherenkov muon spectrometer uses an absolute momentum measurement resolution, $\sigma_p = \pm 0.5$ GeV/c, and uniform size for all radiator chambers.

In this work, we explore a C$_3$F$_8$-based non-linear Cherenkov muon spectrometer which uses variable increasing lengths for radiators. This allows the threshold muon momentum level to increases and to balance the number of Cherenkov photons in all radiators. In addition, the momentum measurement uncertainty becomes stable for all momentum levels by using a relative resolution, $\sigma_p/p$, instead of the absolute resolution. Although the relative resolution slightly varies depending on each muon momentum level, the average resolution is 18%. In the end, we show that our proposed Cherenkov muon spectrometer has high classification rate (CR) for all momentum levels (mean CR = 90.08%) and the cosmic ray muon spectrum was successfully reconstructed using five threshold momentum levels, $p_{th}$ = 0.5, 1.0, 2.0, 4.0, and 8.0 GeV/c, with a relative resolution, $\sigma_p/p$ = 18%, an improvement by a factor of 2 compared to previous designs.

## GAS CHERENKOV RADIATORS

In particle physics, the Cherenkov radiation is often used to detect and identify charged particles that move faster than the speed of light in an optically transparent medium, or a Cherenkov radiator. Although the state of Cherenkov radiator can be solid or liquid, gas radiators are commonly exploited, especially in Cherenkov threshold detectors, since the threshold condition for the Cherenkov effect can vary upon its pressure. The relationship between gas pressure, $p$, and the index of refraction of gas, $n$, is given by the Lorentz-Lorenz equation [15]:

$$n^2 - 1 \approx \frac{3A_m p}{RT} \quad (1)$$

where $A_m$ is the molecular refractivity, $R$ is the universal gas constant, and $T$ is the temperature in K. The required kinetic conditions to induce Cherenkov radiation in the radiator is determined by the index of refraction. The threshold momentum, $p_{th}$, for a charged particle in the medium is given by:

$$p_{th}c = \frac{mc^2}{\sqrt{n^2 - 1}} \quad (2)$$

where $c$ is the speed of light, $m$ is the rest mass of particle. Hence, a new relationship between gas pressure or temperature and the index of refraction is derived by substituting Eq (1) to (2):

$$p_{th}c = mc^2 \left[\frac{3A_m p}{RT}\right]^{-1/2} \quad (3)$$

As candidates for Cherenkov radiator, CO$_2$, R1234yf, C$_4$F$_{10}$, and C$_3$F$_8$, are studied in this work. C$_4$F$_{10}$ and CO$_2$ are gas radiators for the Cherenkov threshold detectors in the Jefferson Lab[16] and CERN[17], respectively. The R1234yf is a promising gas radiators that will replace R12 to reduce the Ozone Depletion Potential (ODP) level[18]. The properties for four Cherenkov gas radiators are summarized in TABLE 1. The variance of both threshold momenta for muons ($m = m_\mu$) and indices of refraction for four gas radiators as a function of pressure (20 to 800 kPa) and temperature (273 to 373 K) are shown in Fig. 1.

TABLE I. Gas Properties for Four Gas Cherenkov Radiator Candidates at 25°C

|  | $CO_2$ | R1234yf | $C_4F_{10}$ | $C_3F_8$ |
|---|---|---|---|---|
| Vapor pressure [MPa] | 5.7 | 0.673 | 0.380 | 0.820 |
| Vapor density [kg/m$^3$] | 1.977 | 37.6 | 24.6 | 12.5 |
| Molecular weight [g/mol] | 44.01 | 114.04 | 236.03 | 188.02 |
| Refractive index [−] | 1.00045 | 1.0010 | 1.0015 | 1.0011 |
| Relative Cherenkov photon yield | 0.33 | 0.74 | 1 | 0.81 |
| Polarizability, $\alpha$ [× $10^{-30}$ m$^3$] | 2.59 | 6.2 | 8.44 | 7.4 |

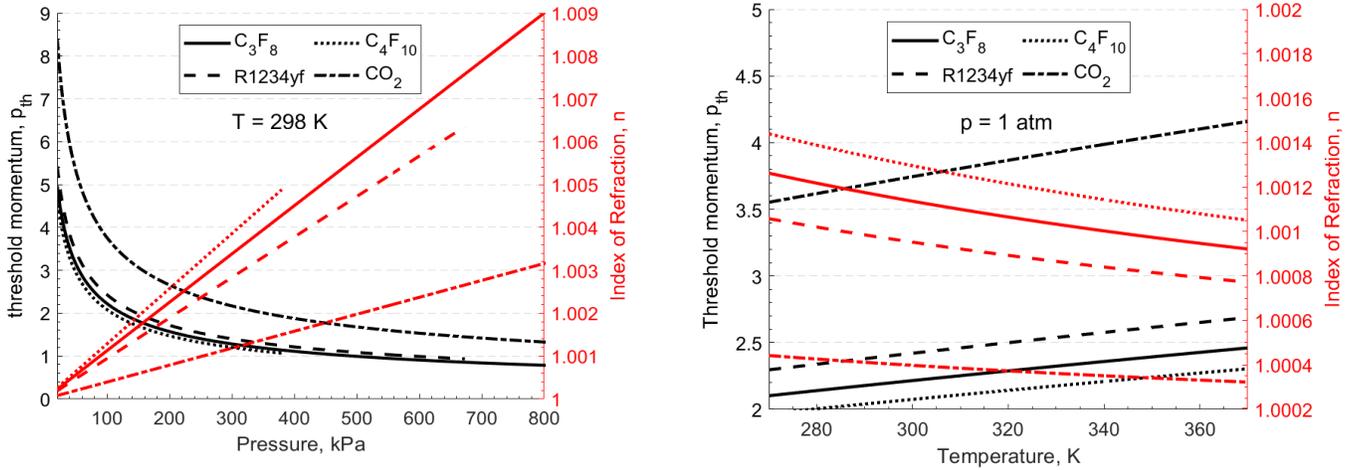

Fig. 1. Variation of the Cherenkov threshold momenta and indices of refraction for four gas radiators, $C_3F_8$, R1234yf, $C_4F_{10}$, and $CO_2$, as a function of pressure (left) and temperature (right). It is noted that the graphs for $C_4F_{10}$ and R1234yf are discontinued due to their pressure limits, 380 and 673 kPa, respectively.

## PRINCIPLE OF CHERENKOV MUON SPECTROMETER

The index of refraction and Cherenkov threshold momentum for gas radiators can be changed by pressurizing the gas radiator without the need to replace any materials. Cooling or heating has much less effect in terms of varying the index of refraction of gas (see Fig. 1). We carefully calibrated the pressure of gas radiators to yield necessary Cherenkov threshold levels to measure muon momentum. When a muon passes multiple pressurized gas radiators that are placed on the muon traveling path, none to all radiators emit Cherenkov radiation depending on their index of refraction and muon momentum. By analyzing the signals from all radiators, we are able to estimate the actual muon momentum. For instance, a schematic of momentum measurement process using the Cherenkov muon spectrometer when the actual muon momentum is 3.1 GeV/c is illustrated in Fig. 2.

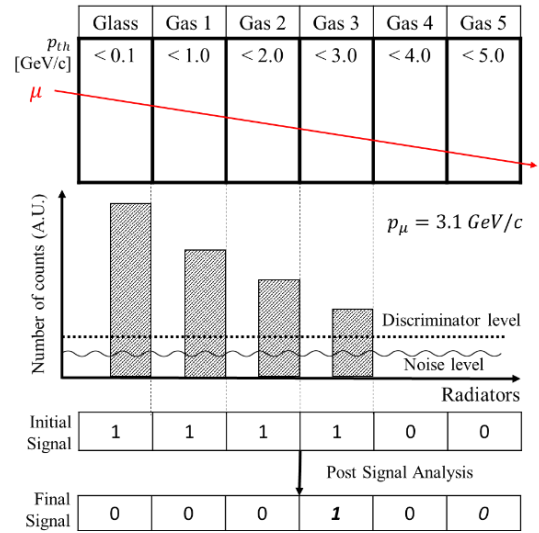

Fig. 2. A schematic of momentum measurement process using Cherenkov muon spectrometer when $p_\mu$ = 3.1 GeV/c. It is noted that the final signal correctly indicates the actual muon momentum range, $3 < p_\mu < 4$ GeV/c [19].

## RESULTS

In this work, our proposed Cherenkov muon spectrometer has five non-linear muon threshold momentum levels, 0.5, 1.0, 2.0, 4.0, and 8.0 GeV/c that consist of one silica aerogel ($SiO_2$) and four pressurized $C_3F_8$ radiators, respectively. We use non-linear threshold momentum levels to provide a stable relative momentum measurement resolution, $\sigma_p/p$, instead of an absolute measurement resolution, $\sigma_p$. In addition, we use diverse muon path length, or radiator size, to balance the overall Cherenkov photon yields in all radiators. The benefits of non-linear and diverse design are: (i) a balanced Cherenkov photon signal, especially at high momentum levels, and (ii) a stable momentum measurement uncertainty. The characteristics of two designs, linear and non-linear, are compared in TABLE 2. A schematic of non-linear Cherenkov muon spectrometer with properties used in Geant4 simulations and an example of optical photon emission when $p_\mu$ = 10 GeV/c are shown in Fig. 3.

When muons travel in the medium, the emission of scintillation and transition radiation are expected in addition to the Cherenkov radiation. In this work, photon signals by them are considered as noise. To quantify the accuracy of muon momentum measurement, we introduced a classification rate (CR) which is a probability that the actual muon momentum is correctly classified in the system. We also introduced a logical signal discriminator which uniformly deducts 1, 2, or 3 from the final signals to eliminate the predictable signals from the scintillation and transition radiation. The results of CR as a function of muon momentum with various discriminator levels are shown in Fig. 4. We demonstrated that CR can be more stable and improved by applying a combination of various discriminator levels ("2" for < 8 GeV/c and "0" for > 8 GeV/c), then the mean CR is 90.08%.

To demonstrate its performance, the cosmic ray muon spectrum was reconstructed using five non-linear muon threshold momentum levels, 0.5, 1.0, 2.0, 4.0, and 8.0 GeV/c and the results were compared with the simulated spectrum using "Muon_generator_v3" which was developed based on semi-empirical models in Fig. 5[20,21]. The reconstructed cosmic ray muon spectrum makes a good agreement with the actual muon spectrum within a 1σ error in most momentum levels.

TABLE II. Comparison of Two Cherenkov Muon Spectrometer Designs

|  | **Linear** | **Nonlinear** |
|---|---|---|
| Radiator size | Uniform | Diverse |
| Threshold momentum levels | linear | nonlinear |
| Material | $CO_2$ | $C_3F_8$ |
| Momentum resolution | $\sigma_p = \pm 0.5$ GeV/c $\sigma_p/p \mid_{avg} = 0.36$ | $\sigma_p/p \mid_{avg} = 0.18$ |

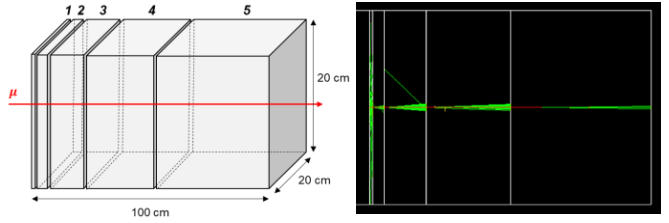

|  | *1* | *2* | *3* | *4* | *5* |
|---|---|---|---|---|---|
| Material | $SiO_2$ | $C_3F_8$ | $C_3F_8$ | $C_3F_8$ | $C_3F_8$ |
| L [cm] | 1 | 4 | 15 | 30 | 50 |
| $p_{th}$ [GeV/c] | 0.5 | 1.0 | 2.0 | 4.0 | 8.0 |
| n-1 | 2.2E-2 | 5.6E-3 | 1.4E-3 | 3.5E-4 | 8.7E-5 |
| P [kPa] | - | 579.28 | 144.82 | 36.21 | 9.05 |
| ρ [kg/m3] | 2270 | 50 | 10 | 2.5 | 0.6 |

Fig. 3. A schematic of non-linear Cherenkov muon spectrometer with the numbers which represent the radiator IDs (left) and visualized Geant4 simulation when $p_\mu$ = 10 GeV/c (right). Red and green represent a negative muon and optical photons, respectively. The material properties and dimensions for all radiators (1 – 5) used in Geant4 simulations are summarized in the table.

## CONCLUSION

A new Cherenkov muon spectrometer using diversely sized $C_3F_8$ gas radiators and nonlinear threshold muon momentum levels was studied using Geant4 simulations. Despite its compact size (<1m$^3$), the performance of the spectrometer was significantly improved over the earlier design (linear and uniform) in two aspects, (i) a stable momentum measurement resolution and (ii) a balanced expected Cherenkov photon yields in all radiators. In addition, $CO_2$ gas radiators are replaced by heavier gas radiators, $C_3F_8$, and the sizes of radiators for high momentum levels (low gas pressure) are enlarged to increase the Cherenkov photon yields. Specifically, the absolute momentum resolution, $\sigma_p$, for all levels is replaced by the relative momentum resolution, $\sigma_p/p$.

To demonstrate the functionality of our proposed Cherenkov muon spectrometer, we presented the results of classification rates as a function of muon momentum with various discriminator levels and reconstructed cosmic ray muon spectrum. When the combination of two or more discriminators are used, the average classification rate is 90.08%. In addition, our proposed Cherenkov muon spectrometer successfully reconstructed the cosmic ray muon spectrum and the results show a good agreement with the actual spectrum within 1σ in most momentum levels.

## ACKNOWLEDGEMENTS

This research is being performed using funding from the Purdue College of Engineering and the School of Nuclear Engineering.

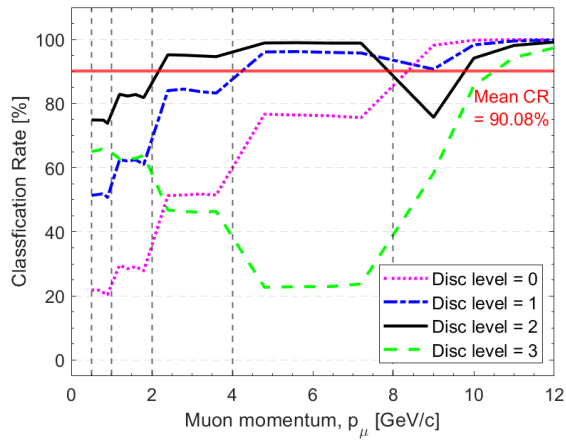 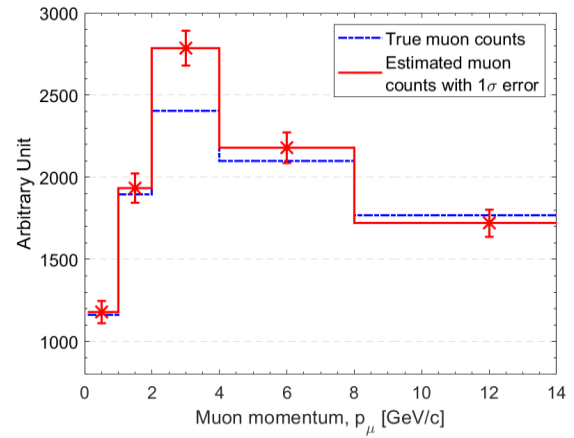

Fig. 4. Classification rates as a function of muon momentum (0.5 to 12.0 GeV/c) using various discriminator levels for $C_3F_8$ nonlinear muon spectrometer. The vertical dashed lines represent threshold momentum levels.

Fig. 5. Reconstructed cosmic ray muon spectrum (0 to 14 GeV/c) using $C_3F_8$ nonlinear muon spectrometer (solid) and actual spectrum (dotted). Error bars represents 1σ and the number of muons used in the simulations is $10^4$.

**REFERENCES**

1. M. ÖSTERLUND et al., "Tomography of canisters for spent nuclear fuel," Proc. Sci., 3 (2006).
2. S. CHATZIDAKIS, C. K. CHOI, and L. H. TSOUKALAS, "Analysis of Spent Nuclear Fuel Imaging Using Multiple Coulomb Scattering of Cosmic Muons," IEEE Trans. Nucl. Sci. **63** 6, 2866, IEEE (2016).
3. A. ERLANDSON et al., "One-Sided Muon Tomography — A Portable Method for Imaging Critical Infrastructure with a Single Muon Detector," CNL Nucl. Rev. **7** 1, 1 (2018).
4. J. PERRY et al., "Imaging a nuclear reactor using cosmic ray muons," J. Appl. Phys. **113** 184909 (2013).
5. V. ANGHEL et al., "Cosmic Ray Muon Tomography System Using Drift Chambers for the Detection of Special Nuclear Materials," IEEE NSS MIC, 547, IEEE (2010).
6. J. BAE and S. CHATZIDAKIS, "The Effect of Cosmic Ray Muon Momentum Measurement for Monitoring Shielded Special Nuclear Materials," INMM/ESARDA Jt. Annu. Meet. (2021).
7. E. V BUGAEV et al., "Atmospheric muon flux at sea level, underground, and underwater," Phys. Rev. D **58** (1998).
8. T. SATO, "Analytical model for estimating the zenith angle dependence of terrestrial cosmic ray fluxes," PLoS One **11** 8, 1 (2016); https://doi.org/10.1371/journal.pone.0160390.
9. J. BAE and S. CHATZIDAKIS, "A New Semi-Empirical Model for Cosmic Ray Muon Flux Estimation," Prog. Theor. Exp. Phys. **ptac016** (2022).
10. J. BAE, S. CHATZIDAKIS, and R. BEAN, "Effective Solid Angle Model and Monte Carlo Method: Improved Estimations to Measure Cosmic Muon Intensity at Sea Level in All Zenith Angles," Int. Conf. Nucl. Eng. (2021).
11. K. HAGIWARA et al., "Review of Particle Properties (Particle Data Group)," Phys. Rev. D **66** (2002).
12. K. N. BOROZDIN et al., "Radiographic imaging with cosmic-ray muons," Nature **422** 277, 20 (2003).
13. J. BAE and S. CHATZIDAKIS, "A Cosmic Ray Muon Spectrometer Using Pressurized Gaseous Cherenkov Radiators," IEEE Nucl. Sci. Symp. Med. Imaging Conf. (2021).
14. J. BAE and S. CHATZIDAKIS, "Fieldable Muon Momentum Measurement using Coupled Pressurized Gaseous Cherenkov Detectors," Trans. Am. Nucl. Soc. **125** 1, 400 (2021).
15. H. KRAGH, "The Lorenz-Lorentz Formula : Origin and Early History," Substantia **2**, 7 (2018).
16. G. NEIL, "Thomas Jefferson National Accelerator Facility FEL Industrial Applications," in Joint Accelerator Conferences, pp. 88–92 (1998).
17. H. W. ATHERTON et al., "Precise measurements of particle production by 400 GeV/c protons on beryllium targets" (1980).
18. A. H. HARVEY, E. PAULECHKA, and P. F. EGAN, "Candidates to replace R-12 as a radiator gas in Cherenkov detectors," Nucl. Instruments Methods Phys. Res. Sect. B **425**, 38 (2018).
19. J. BAE and S. CHATZIDAKIS, "Fieldable muon spectrometer using multi-layer pressurized gas Cherenkov radiators and its applications," Sci. Rep. (2022).
20. S. CHATZIDAKIS and L. H. TSOUKALAS, "A Geant4-MATLAB Muon Generator for Monte-Carlo Simulations," 4 (2016); https://doi.org/10.13140/RG.2.2.31871.41128.
21. S. CHATZIDAKIS, "Muon Generator" (2021).